\documentclass[aps,twocolumn,nofootinbib,groupedaddress,amsfonts,floatfix]{revtex4-1} 
\pdfoutput=1 
\usepackage{graphicx,amsmath,amssymb,amstext}
\usepackage{amssymb,amsbsy,amsfonts,amsthm,color} 

\usepackage{epsfig}
\usepackage{subfigure} 
\usepackage{hyperref}
\usepackage{cancel}
\usepackage{scrextend}
\usepackage{tikz}
\usepackage{tikz-cd}
\usepackage{graphicx}

\graphicspath{{Figures/}}

\usepackage{color}

\usetikzlibrary{shapes.arrows, fadings}
\usetikzlibrary{positioning,arrows}
\usetikzlibrary{decorations.pathmorphing}
\usetikzlibrary{decorations.markings}
\usetikzlibrary{shapes.symbols}
\usetikzlibrary{shapes.misc}
\usetikzlibrary{backgrounds}

\tikzfading[name=fade right,
  left color=transparent!0, right color=transparent!100]
\tikzfading[name=fade left,
  right color=transparent!0, left color=transparent!100]

\tikzfading[name=fade left,
  left color=transparent!100, right color=transparent!0]

\def\beq{\begin{equation}}
\def\eeq{\end{equation}}
\newcommand{\Beq}{\begin{eqnarray}}
\newcommand{\Eeq}{\end{eqnarray}}

\def\lsim{\mathrel {\vcenter {\baselineskip 0pt \kern 0pt \hbox{$<$} \kern 0pt \hbox{$\sim$} }}}

\def\gsim{\mathrel {\vcenter {\baselineskip 0pt \kern 0pt \hbox{$>$} \kern 0pt \hbox{$\sim$} }}}

\begin{document}

\title{Prethermalization Production of Dark Matter}

\author{Marcos A.~G.~Garcia and Mustafa~A.~Amin}
\email{marcos.garcia@rice.edu, mustafa.a.amin@rice.edu}
\affiliation{Department of Physics and Astronomy, Rice University, Houston, Texas 77005-1827, USA}
\begin{abstract}

At the end of inflation, the inflaton field decays into an initially nonthermal population of relativistic particles which eventually thermalize. We consider the production of dark matter via freeze-in from this relativistic plasma, focusing on the prethermal phase.
We find that for a production cross section $\sigma(E)\sim E^n$ with $n> 2$, the present dark matter abundance is produced during the prethermal phase of its progenitors. For $n\le 2$, entropy production during reheating makes the nonthermal contribution to the present dark matter abundance subdominant compared to that produced thermally. As specific examples, we verify that the nonthermal contribution is irrelevant for gravitino production in low scale supersymmetric models ($n=0$) and is dominant for gravitino production in high scale supersymmetry models ($n=6$).



\end{abstract} 
\maketitle

\section{Introduction} \label{sect:intro}

The nonthermal universe at the end of inflation is often assumed to be hidden from observations due to subsequent thermalization and radiation domination of the universe prior to big bang nucleosynthesis (BBN).\footnote{Herein thermalization stands for the local kinetic and chemical equilibration of the inflation decay products. It is not to be confused with the onset of radiation domination, which may occur before or after thermal equilibrium is reached.} However, gravitational effects and stable nongravitational relics, if produced in this period, are sensitive to/provide potential probes of this era. Examples of gravitational effects include production of gravitational waves, changes in the expansion history of the universe, and production of primordial black holes. Nongravitational relics include the matter-antimatter asymmetry and the dark matter (DM) abundance
~\cite{reviews1,*reviews7,*reviews8,*therm1,*therm10}.  In this paper, we provide conditions under which the present DM abundance can be sensitive to the nonthermal phase immediately following inflation, and carry out a detailed calculation of this abundance. More specifically, we study concrete DM models where nonthermal conditions change the predictions from the thermal scenario by many orders of magnitude.

We consider the case where DM is primarily produced via {\it freeze-in} from a relativistic plasma, which in turn is produced via a small coupling to the inflaton.\footnote{We do not consider the case of DM produced nonthermally through direct decays; this scenario is studied in~\cite{Hall:2009bx,EGNOP,Blinov:2014nla,Kim:2016spf}.} In this paper 
we restrict ourselves to perturbative processes during reheating after inflation. Under these conditions we arrive at a rather simple conclusion:\\

\begin{addmargin}[1.5em]{1.5em}
If the total cross section for DM production from the relativistic plasma has a sufficiently steep dependence on energy: $\sigma(E)\propto E^n$ with integer $n>2$, and the thermalization of the inflaton decay products is mediated by the Standard Model (SM) gauge interactions, then it is the production of DM from the nonthermal phase of the relativistic plasma, at the earliest stages after the end of inflation, that predominantly determines the DM abundance. \\
\end{addmargin}

For gauge-mediated thermalization, our results supersede other ultraviolet (UV) sensitive DM production scenarios considered in the literature. In~\cite{Garcia:2017tuj}, one of us found that the dependence of the DM abundance on the highest temperature during reheating $T_{\rm max}$  is obtained for $n\ge 6$, assuming a relativistic plasma in thermal equilibrium. \footnote{Note that by thermal/nonthermal we are referring to whether degrees of freedom in the relativistic plasma have a momentum distribution given by the chemical equilibrium ones (Fermi-Dirac/Bose-Einstein). }. In the present paper, by including nonthermal effects, we find UV sensitivity at the milder power $n>2$, which makes our result relevant for a broader range of mo\-dels~\cite{bcdm,DMO,Dudas:2018npp,Dudas:2017kfz,Bernal:2018qlk,Bhattacharyya:2018evo}.

The rest of the paper is organized as follows. We briefly review the DM production mechanisms in the literature in Sec.~\ref{sec:review}. Therein we also introduce relevant notation and definitions, and we schematically summarize our results. We carry out the detailed calculation of nonthermal DM production in Sec.~\ref{sec:preth}. We then specialize to the application of the formalism to gravitino production in Sec.~\ref{sec:apps}, in the case of weak scale supersymmetry breaking ($n=0$), as well as that of very high energy supersymmetry breaking models, where the population of almost the whole spectrum of superpartners is kinematically forbidden during reheating ($n=6$). Section~\ref{sec:conc} is reserved for our summary and future applications.

\subsection{Dark matter production scenarios (review)}
\label{sec:review}


Particle DM production scenarios can be roughly divided into two classes.\footnote{This assumes implicitly the absence of an independently thermalized dark sector~\cite{Bernal:2017kxu,Adshead:2016xxj}}  In thermal {\it freeze-out} models, the DM particles (hereafter denoted by $\chi$) interact sufficiently strongly at high energies with the primordial relativistic plasma (denoted by $\gamma$) keeping $\gamma$ and $\chi$ in thermal equilibrium. Once $\chi$ becomes nonrelativistic and the expansion rate overtakes the annihilation rate, the DM yield\footnote{We denote by $\gamma$ a single degree of freedom of the relativistic plasma; for a species in thermal equilibrium, $n_{\gamma} \rightarrow n_{\gamma}^{\rm T} =\frac{\zeta(3)}{\pi^2}T^3$.}\footnote{In the absence of entropy conservation, this ratio is better suited for calculations than the DM-to-entropy ratio~\cite{EGNOP}.}
 \Beq
 Y_\chi\equiv \frac{n_\chi}{n_\gamma}
 \Eeq
freezes-out of equilibrium. $Y_{\chi}$ is therefore insensitive to the early thermal history of the universe up to the decoupling temperature, $T_{\rm dec}\lesssim 20\,m_{\rm DM}$ for processes mediated by the weak interaction~\cite{kolb1990the}. That is, freeze-out is infrared (IR) dominated with respect to the thermal history that follows after reheating, $T_{\rm dec}\ll T_{\rm reh}$. 

In contrast, {\em freeze-in} scenarios assume that the coupling of DM with the relativistic plasma is so suppressed that DM never reaches thermal equilibrium with the plasma~\cite{McDonald:2001vt,Hall:2009bx,Elahi:2014fsa}. In this case then, the primordial abundance is determined by forward processes (plasma$\rightarrow$DM) rather than the annihilation rate, and can only be significantly reduced by DM decay. Moreover, freeze-in can be either IR-dominated~\cite{Silveira:1985rk,McDonald:1993ex,Burgess:2000yq,Yaguna:2011qn,Bernal:2015ova,Dodelson:1993je,Shi:1998km,Abazajian:2001nj,Shakya:2015xnx,Adhikari:2016bei}, or UV-dominated, as is the case for DM production through heavy mediators~\cite{moqz,mnoqz,Blennow:2013jba,Chu:2013jja,noz,brian,Chen:2017kvz,Bernal:2018qlk,Elahi:2014fsa} or the thermal production of gravitinos~\cite{ehnos,Ellis:1986zt,Moroi:1994rs,Endo:2006zj,Nakamura:2006uc,Dine:2006ii,Nakayama:2012hy,Jeong:2012en,nos,kl,ekn, Juszkiewicz:gg,mmy,Kawasaki:1994af,Moroi:1995fs,enor,Giudice:1999am,bbb,Pradler:2006qh,ps2,rs,Nilles:2001ry,Nilles:2001fg,EGNOP,Ema:2016oxl,bcdm,DMO,Dudas:2018npp,Dudas:2017kfz,Hook:2018sai}. 

\begin{figure*}[t!]
\begin{center}
\scalebox{1.05}{
\begin{tikzpicture}%
    	\node[anchor=south west,inner sep=0] (image) at (0,0) {\scalebox{0.46}{\includegraphics{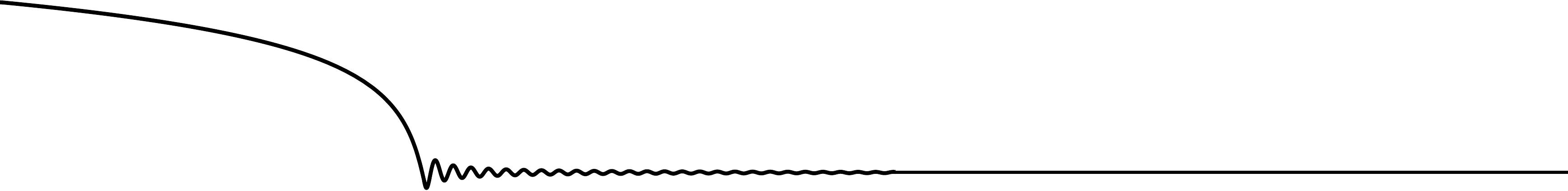}}};%
    	\begin{scope}[x={(image.south east)},y={(image.north west)}]%
	\fill[gray!60,path fading=fade left] (0.263,-0.15) rectangle (0.5725,1.05);
	\fill[gray!60] (0.572,-0.15) rectangle (1,1.05);
	\draw[->,line width=.02cm] (0,-0.15) - - (1,-0.15) ;
	\node at (0.99,-0.37) {$t$};
	\draw[line width=.02cm] (0.263,-0.22) - - (0.263,-0.1) ;
	\node at (0.263,-0.37) {$t_{\rm end}$};
	\draw[line width=.02cm] (0.572,-0.22) - - (0.572,-0.1) ;
	\node at (0.572,-0.37) {$t_{\rm reh}$};
	\draw[line width=.02cm] (0.417,-0.22) - - (0.417,-0.1) ;
	\node at (0.417,-0.37) {$t_{\rm th}$};
	\draw[->,line width=.02cm] (0,-0.15) - - (0,1.25) ;
	\node at (0.04,0.81) {\footnotesize $\phi$};
	\draw[<->,line width=.02cm] (0.263,0.6) - - (0.417,0.6) ;
	\node at (0.34,0.73) {\footnotesize non-thermal $\gamma$};
	\draw[line width=.02cm] (0.417,0.51) - - (0.417,0.69) ;
	\draw[<-,line width=.02cm] (0.417,0.6) - - (1,0.6) ;
	\node at (0.71,0.73) {\footnotesize thermal $\gamma$};
	\draw[<->,line width=.02cm] (0.263,1.05) - - (0.572,1.05) ;
	\node at (0.4175,1.15) {\footnotesize matter domination};
	\draw[line width=.02cm] (0.572,0.96) - - (0.572,1.14) ;
	\draw[<-,line width=.02cm] (0.572,1.05) - - (1,1.05) ;
	\node at (0.786,1.15) {\footnotesize radiation domination};
	\draw[line width=.02cm] (0.263,0.96) - - (0.263,1.14) ;
	\draw[->,line width=.02cm] (0,1.05) - - (0.263,1.05) ;
	\node at (0.132,1.15) {\footnotesize inflation};
	\draw[->,line width=.02cm] (0,-2.6) - - (0,-0.7) ;
	\draw[->,line width=.02cm] (0,-2.6) - - (1,-2.6) ;
	\node at (0.06,-0.74) {\footnotesize $Y_{\chi}=\dfrac{n_{\chi}}{n_{\gamma}}$};
	\node at (0.9,-0.9) {\footnotesize $\sigma_{\gamma\gamma\rightarrow\chi\chi}(E)\propto E^n$};
	\node at (0.99,-2.82) {$t$};
	\draw[line width=.02cm] (0.263,-2.67) - - (0.263,-2.55) ;
	\node at (0.263,-2.82) {$t_{\rm end}$};
	\draw[line width=.02cm] (0.572,-2.67) - - (0.572,-2.55) ;
	\node at (0.572,-2.82) {$t_{\rm reh}$};
	\draw[line width=.02cm] (0.417,-2.67) - - (0.417,-2.55) ;
	\node at (0.417,-2.82) {$t_{\rm th}$};
	\draw[line width=.02cm] (0.84,-2.67) - - (0.84,-2.55) ;
	\node at (0.84,-2.82) {\footnotesize $m_{\chi}\sim T$};
	%
	\node at (0.83,-2.22) {\scalebox{0.30}{\includegraphics{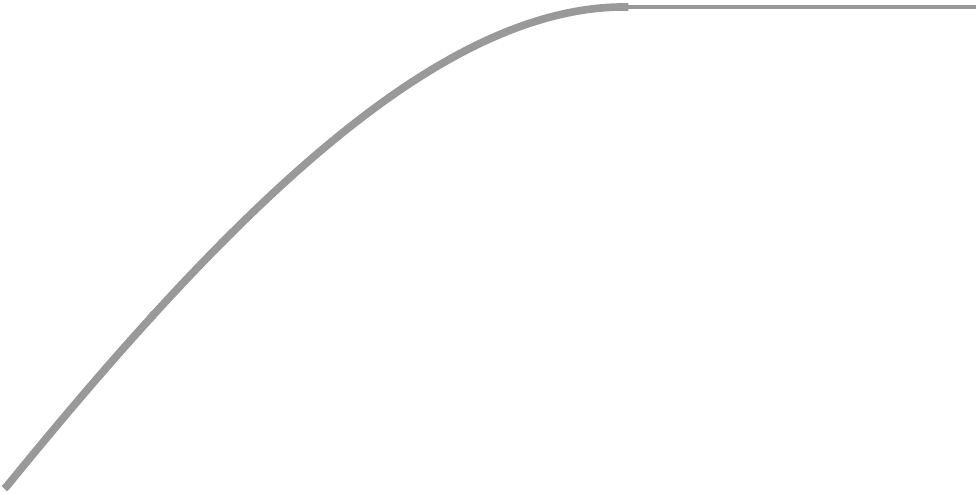}}};
	\node at (0.56,-2.16) {\scalebox{0.30}{\includegraphics{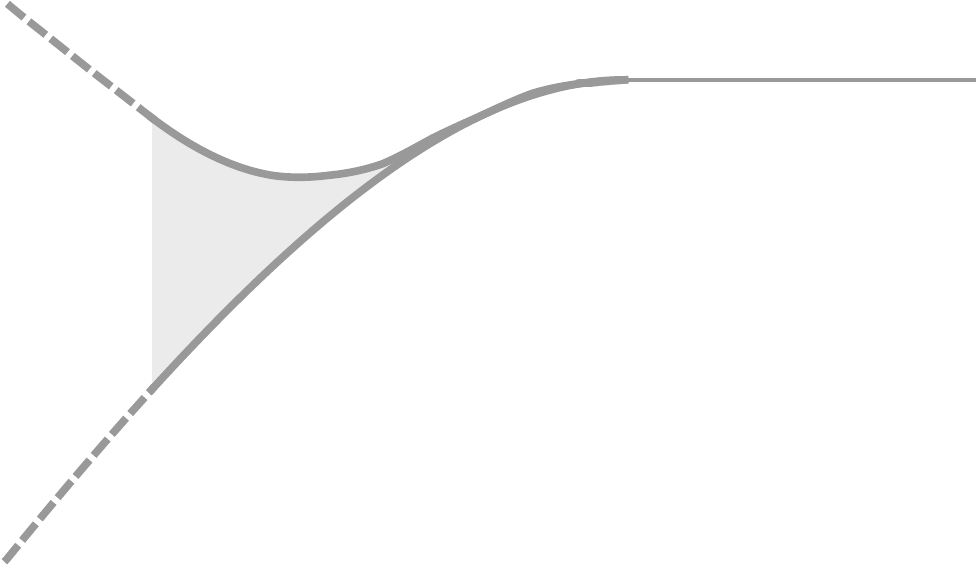}}};
	\node at (0.50,-1.77) {\scalebox{0.378}{\includegraphics{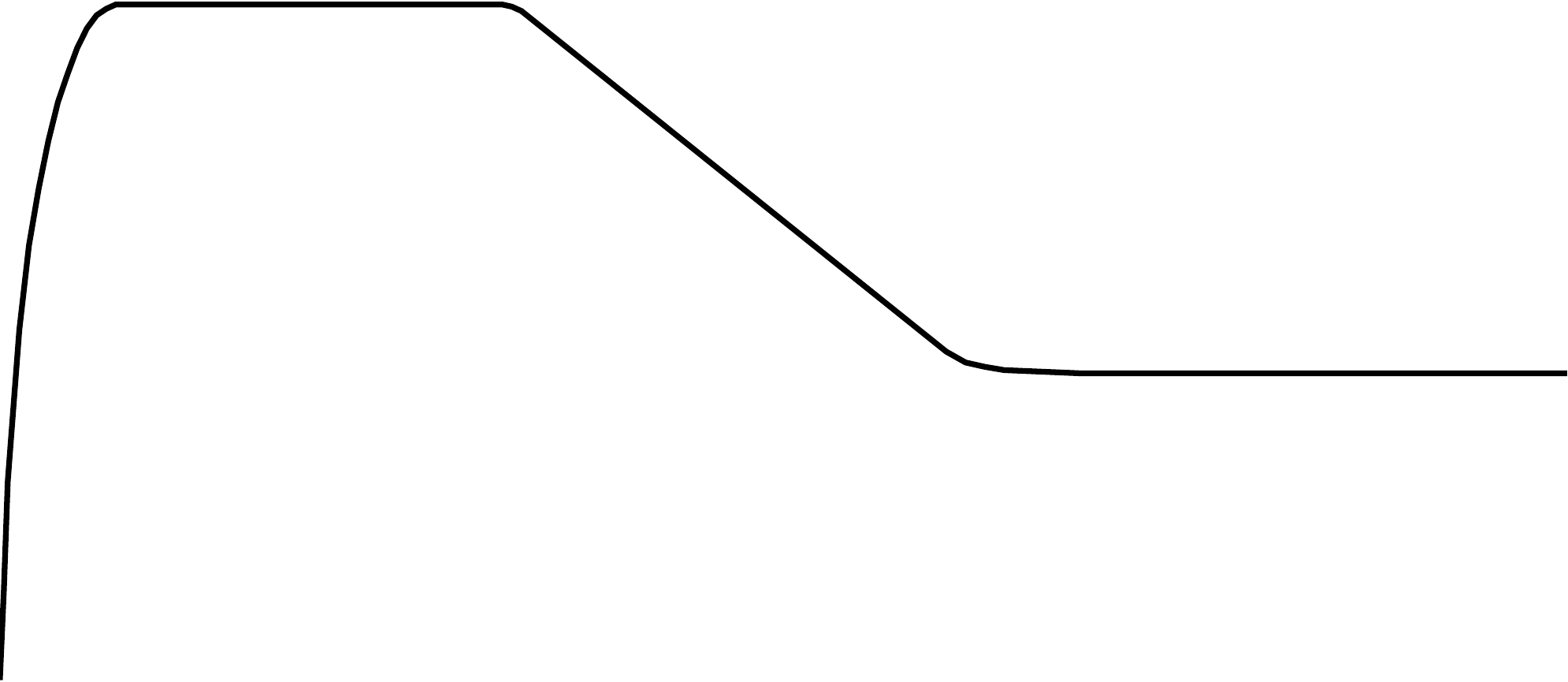}}};
	\fill[white] (0.6,-1.8) rectangle (0.75,-1.9);
	\fill[white] (0.86,-1.8) rectangle (0.97,-1.9);
	\draw[line width=.036cm] (0.6,-1.85) - - (1,-1.85) ;
	%
	\node at (0.83,-2.2) {\footnotesize $n<-1$};
	\node at (0.59,-2.2) {\footnotesize $2\geq n>-1$};
	\node at (0.355,-0.85) {\footnotesize $n>2$};
	\node at (0.95,-1.7) {\footnotesize $Y_{\chi}^{\rm obs}\ll 1$};
	\node at (0.12,-1.55) {\footnotesize non-thermal};
	\node at (0.127,-1.7) {\footnotesize dominated};
	\node at (0.138,-2.0) {\footnotesize thermal};
	\node at (0.127,-2.15) {\footnotesize dominated};
	\draw[->,line width=.02cm] (0.185,-1.64) - - (0.26,-1.64) ;
	\draw[->,line width=.02cm] (0.185,-2.07) - - (0.47,-2.07) ;
	\end{scope}%
\end{tikzpicture}}%
\caption{ Upper panel: The evolution of the inflaton field $\phi$ (black curve) and its relativistic decay products $\gamma$ (gray shading). The time $t_{\rm end}$ denotes the end of inflation, $t_{\rm th}$ is the time of full thermalization of $\gamma$, and $t_{\rm reh}$ denotes the end of reheating when the inflaton decays almost entirely to $\gamma$. After $t_{\rm reh}$, the universe becomes radiation dominated. Dark matter $\chi$ can be produced from $\gamma$ for $t>t_{\rm end}$ in the nonthermal and thermal phases.  Lower panel: Different paths to the observed dark matter abundance. For a production cross section $\sigma(E)\propto E^n$, the production of $\chi$ from thermal $\gamma$  dominates for $2\geq n$ (gray curves), while the production of $\chi$ from nonthermal $\gamma$ dominates for $n>2$ (black curves). The $n>2$ case is sensitive to the earliest stages at the end of inflation.}
\label{fig:introfig}
\end{center}
 \end{figure*}
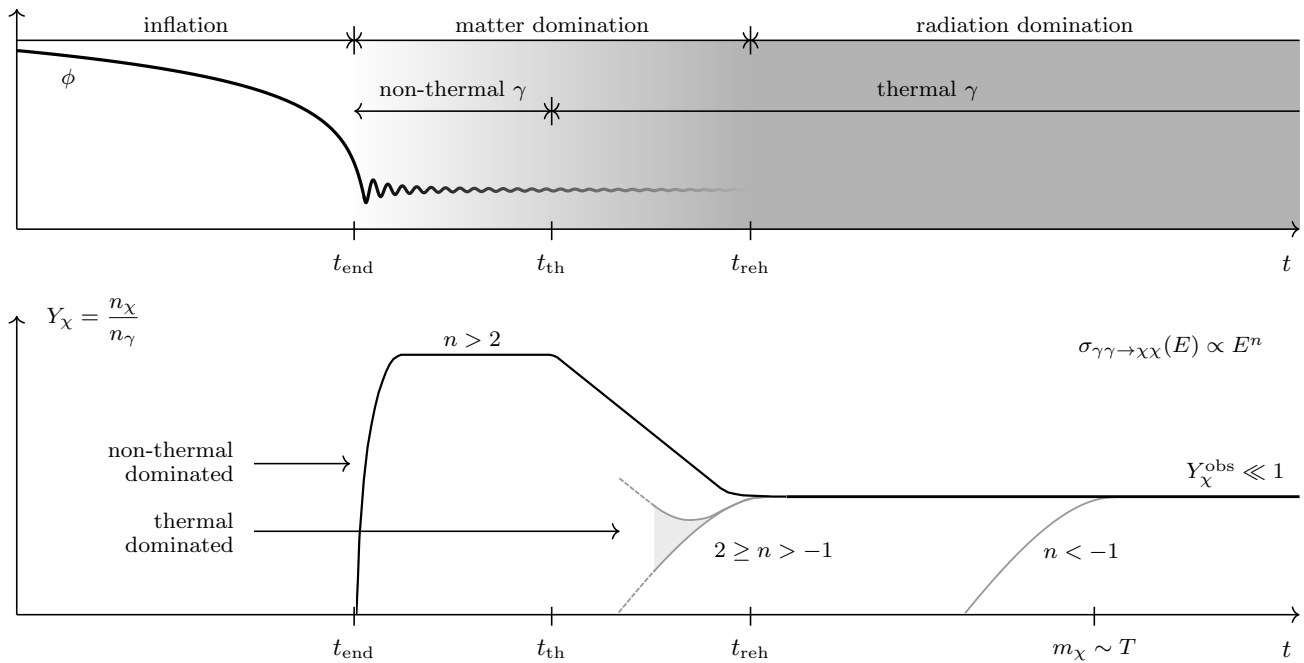

Let us assume that DM production takes place via $2\rightarrow2$ scattering processes in the presence of a thermalized background, with the integrated effective cross section
\beq\label{eq:sigmani}
\sigma(s) \propto \frac{s^{n/2}}{M^{n+2}}\,,
\eeq
where $\sqrt{s}=E$ is the center of mass energy of the scattering. Figure~\ref{fig:introfig} schematically shows freeze-in scenarios for different values of $n$, which we discuss further below. For non-negative $n$, the cross section violates unitarity at very high energies, and we assume that it corresponds to a low-energy effective description of a UV-complete theory. This type of energy dependence will generically arise if the DM sector, decoupled at low energies, communicates to the visible sector via high dimensional operators. The mass scale $M$ in (\ref{eq:sigmani}) can be thought to be parametrically related to the mass of a heavy mediator in the UV theory.


In the absence of entropy production (for example, during the radiation dominated era that follows after the end of inflationary reheating) freeze-in is UV dominated for $n>-1$; that is, $Y_{\chi}(T\ll T_{\rm reh})\sim Y_{\chi}(T_{\rm reh})$~\cite{Elahi:2014fsa,EGNOP,Garcia:2017tuj}. This result then holds under the assumption that the inflaton $\phi$ instantaneously decays into a thermal bath of initial temperature
\beq
T_{\rm reh}\;=\;\left(\frac{40}{\pi^2 g_{\rm reh}}\right)^{1/4}(\Gamma_{\phi}M_P)^{1/2}\,,
\eeq
at the decay time $t_{\rm reh} = \Gamma^{-1}_{\phi}$, where $g_{\rm reh}$ denotes the number of effective degrees of freedom in the thermal bath,  $\Gamma_{\phi}$ is the (perturbative) inflaton decay rate, and $M_P=(8\pi G)^{-1/2}$ is the reduced Planck mass. %

However, the reheating process is not instantaneous, but involves a continuous transferal of inflaton energy density into its relativistic decay products, until the former is depleted around the time when the universe becomes dominated by radiation. Entropy is not conserved during reheating, and DM is continuously produced as the thermal plasma is populated. A more detailed calculation which takes into account the finite duration of the reheating process shows that under the assumption of an {\em instantaneously thermalized} background at the maximum temperature during reheating $T_{\rm max}\sim (m_\phi/\Gamma_\phi)^{1/4}T_{\rm reh}$~\cite{EGNOP,Tmax1_1,Tmax1_2,Tmax2}, the freeze-in DM yield takes the form
\Beq
Y^{\rm T}_\chi \propto T_{\rm reh}^7\left(\frac{T_{\rm max}^{n-6}-T_{\rm reh}^{n-6}}{n-6}\right)\,,
\label{eq:Ytherm}
\Eeq
for $n\geq0$, $n\neq 6$. The special $n=6$ case yields a logarithmic dependence: $Y^{\rm T}_\chi\propto T_{\rm reh}^7\ln(T_{\rm max}/T_{\rm reh})$~\cite{Garcia:2017tuj}. Remarkably, this result implies that for $n\geq 6$ the temperature dependence of the DM yield is dominated by $T_{\rm max}$, that is, DM production is {\em UV-dominated during reheating}. Only such a steep temperature dependence of the scattering cross section can be sufficiently competitive with respect to the dilution rate due to entropy production.

It is worth noting that in UV-dominated models, the decay of the inflaton determines the DM yield. In the present epoch, $T\ll T_{\rm reh}$, the DM is nonrelativistic, and its relic abundance will be given by 
\beq\label{eq:omegachi}
\Omega_{\chi} \;= \; \frac{g(T)}{g_{\rm reh}} \frac{m_{\chi}n_{\gamma}Y_{\chi}}{2\rho_c }\,.
\eeq
Accounting only for Standard Model degrees of freedom, $g_{\rm reh}=427/4$ and $g(T\ll 1\,{\rm MeV}) = 3.91$, and a DM mass $m_{\chi}=100\,{\rm GeV}$, the observed DM abundance $\Omega_{\chi}h^2 \simeq 0.12$~\cite{Patrignani:2016xqp} requires the yield at the end of reheating to be $Y_{\chi}\simeq 10^{-9}\ll 1$. The smallness of $Y_{\chi}$ will constrain us to focus on Planck-suppressed decay rates or smaller for $\phi$, $\Gamma_{\phi}<m_{\phi}^3/M_P^2$.

\subsection{Nonthermal effects in dark matter production}
\label{sec:review3}

The high-energy sensitivity discussed above invites us to consider nonthermal effects. Recall that when an inflaton decays, it produces high energy quanta with energy $\sim m_\phi > T_{\rm max}$. Until elastic and inelastic processes within the plasma become sufficiently efficient to maintain kinetic and chemical equilibrium, these not-yet-thermalized decay products can then copiously produce DM particles if the production cross section is UV sensitive. Therefore, if thermalization is significantly delayed compared to the timescale attaining $T_{\rm max}$, the bulk of the DM abundance may be produced {\em before} the onset of thermal equilibrium in the universe. As we will show later, in scenarios for which thermal equilibrium is reached via gauge interactions, this delay is sufficient to lead to a nonthermally produced DM abundance for $n>2$.

A schematic summary of our results is shown in Fig.~\ref{fig:introfig}. The upper panel shows the typical evolution of the inflaton field and its decay products (shown as a shaded background), starting from the slow roll of $\phi$ in the inflationary era, continuing through the reheating epoch, dominated by the pressureless oscillation of $\phi$, and finishing in the radiation dominated era. Note the finite nonthermal  epoch, spanning times between the end of inflation at $t_{\rm end}$ and the time of formation of a fully thermalized background at $t_{\rm th}$. The lower panel shows the evolution of the DM yield in freeze-in models as a function of $n$ in (\ref{eq:sigmani}) For $n<-1$ the cross section favors lower energies, and the bulk of the DM is produced around the time it becomes nonrelativistic. For $2\geq n>-1$, the production is enhanced at higher energies, but this enhancement cannot compete against the dilution due to the decay of the inflaton, and as a result the population created around $t_{\rm reh}$ dominates. For $n>2$, the DM population produced by the nonthermal particles survives after the end of reheating and constitutes the majority of the DM particles.

It is worth emphasizing here that our results are valid for high reheating temperatures, $T_{\rm reh}\gtrsim m_{\chi}$. If this assumption is not realized, freeze-out(in) occurs during reheating and the condition $n>2$ is not necessary for a nonthermally dominated DM abundance. In the scenario of~\cite{Harigaya:2014waa}, for example, $\sigma\sim s^{-1}$ and $T_{\rm reh}\ll m_{\chi}$. DM is produced nonthermally around $t\sim t_{\rm reh}$, via bremsstrahlung of inflaton decay products with energy $\sim m_{\phi}$ in the soft thermal background.  A nonthermal DM abundance is also a feature of scalar DM that can exist in a coherent state at early times, such as very light bosons (see~\cite{Hui:2016ltb,*Marsh:2015xka,*Hlozek:2014lca} for reviews), or in some Higgs-portal models~\cite{Nurmi:2015ema}.

We next consider the prethermalization epoch in detail. The treatment is mostly analytic. Numerical calculations for specific models are done in Sec.~\ref{sec:apps}.

\section{Prethermalization particle production}
\label{sec:preth}

The transfer of the energy density stored in the inflaton field to its decay products during perturbative reheating can be approximated by the following set of equations:
\begin{align}\label{eq:rhoeq1}
\dot{\rho}_{\phi} + 3H\rho_{\phi} + \Gamma_{\phi}\rho_{\phi} &=0\,,\\
\dot{\rho}_{\gamma} + 4H\rho_{\gamma}  -\Gamma_{\phi}\rho_{\phi} &=0\,,\\
\rho_{\phi}+\rho_{\gamma}&=3M_P^2H^2\,,
\end{align}
where we denote by $\rho_{\phi}$ and $\rho_{\gamma}$ the energy densities of the inflaton condensate and that of its relativistic decay products, respectively. Since we are interested in a freeze-in scenario, we implicitly assume that the contribution of DM particles to $\rho_{\gamma}$ is negligible. We also assume that no direct decay channel of the inflaton to DM is available.

\subsection{Before thermalization of $\gamma$}

In the very earliest stages of reheating, $\Gamma_{\phi}(t-t_{\rm end})\ll 1$ where the subscript ``end'' denotes the end of inflation, the thermalization of the inflaton decay products will at best be incomplete. In the limit in which the created particles are not interacting, we can then approximate their number density at a given time by counting the number of inflaton quanta that have decayed up to that instant. Straightforward integration of (\ref{eq:rhoeq1}) leads to 
\begin{align}\label{eq:nphi}
n_{\phi} \;&=\; \frac{\rho_{\phi}}{m_{\phi}} \;=\; \frac{\rho_{\rm end}}{m_{\phi}}\left(\frac{a}{a_{\rm end}}\right)^{-3}e^{-v}\,,\\ \label{eq:ngamma}
n_{\gamma} \;&\sim\;   \frac{\rho_{\rm end}}{m_{\phi}}\left(\frac{a}{a_{\rm end}}\right)^{-3}\left(1-e^{-v}\right)\,,
\end{align}
where $a$ denotes the scale factor, $m_{\phi}$ is the inflaton mass, and
\beq
v\;\equiv\; \Gamma_{\phi}(t-t_{\rm end})
\eeq
is a natural dimensionless time variable, $v_{\rm reh}=1$. 

Let us parametrize the inflaton decay rate as
\beq
\Gamma_{\phi}=\frac{|y|^2}{8\pi}m_{\phi}
\eeq
and denote by $\tilde{g}$ the effective number of relativistic degrees of freedom contributing to the number density of the thermal plasma. It is then easy to verify that the noninteracting decay products are in a regime of ‘underoccupation’ with respect to their thermal counterparts, $\tilde{g}\, n_{\gamma}^{\rm T}/n_{\gamma}>1$, for $|y|\lesssim\mathcal{O}(10^{-5})$, until number-increasing processes become efficient~\cite{Harigaya:2013vwa,EGNOP}. Here $n_{\gamma}^{\rm T}=\zeta(3)T^3/\pi^2$. This ``weak reheating'' regime will be generically realized for Planck-suppressed inflaton decays, and, as we will discuss in the following sections, it corresponds to the regime allowed by gravitino production.

In order to compute the DM production rate we need not only (\ref{eq:ngamma}) but also the phase space distribution function $f_{\gamma}(p,t)$ of the decay products. It is illustrative to derive it directly from the Boltzmann equation. In a comoving frame, the inflaton particles sit in a condensate at rest, with a distribution given by $f_{\phi}=(2\pi)^3n_{\phi}\delta^{3}({\bf p})$. The resulting Boltzmann transport equation for the process $\phi\rightarrow \gamma + \gamma$, assuming a two body decay for simplicity, is then given by
\begin{equation}
\frac{\partial f_{\gamma}}{\partial t} - Hp\frac{\partial f_{\gamma}}{\partial p} = \frac{2\pi^2}{p^2}n_{\phi}\Gamma_{\phi}\delta(p-m_{\phi}/2)\,.
\end{equation}
At very early times the exponential in (\ref{eq:nphi}) may be approximated as a constant; for Planck-suppressed decays and thermalization mediated by SM gauge interactions this approximation holds beyond the equilibration of the relativistic plasma [see Eq.~(\ref{eq:vth}) and below]. The transport equation can then be integrated to yield
\begin{align}\notag
f_{\gamma} \;&=\; \frac{16\pi^2\Gamma_{\phi}}{m_{\phi}^3}\int^t_{t_{\rm end}}\frac{n_{\phi}(t')}{H(t')}\delta(t'-t_0)\,,&& \frac{a(t)}{a(t_0)}=\frac{m_{\phi}}{2p}\\ \label{eq:fgamma}
&\simeq \; 24\pi^2\frac{n_{\gamma}}{m_{\phi}^3}\left(\frac{m_{\phi}}{2p}\right)^{3/2}\theta(m_{\phi}/2-p)\,.
\end{align}
As expected, the energies of these ``hard'' decay products are centered at much higher values, $\langle p\rangle\sim m_{\phi}$, than their would-be thermal counterparts~\cite{McDonald:1999hd,Allahverdi:2000ss}.

Let us consider now the DM creation process $\gamma+\gamma \rightarrow \chi + \chi$. Recalling that the progenitors are underoccupied with respect to their thermal distributions, the Boltzmann equation for the DM density $n_{\chi}$ can simply be written as
\begin{align}\notag
&\dot{n}_{\chi} + 3Hn_{\chi} \;=\; 4g_{\chi}^2g_{\gamma}^2\\
&\quad  \times \int \frac{d^3{\bf k}_1}{(2\pi)^22k_1} \frac{d^3{\bf k}_2}{(2\pi)^22k_2}\,(k_1\cdot k_2) \sigma(s) f_{\gamma}(k_1)f_{\gamma}(k_2)\,,
\end{align}
where $\sigma(s)\equiv\sigma(s)_{\gamma+\gamma \rightarrow \chi+\chi}$, $s$ is the Mandelstam variable, and we have accounted for the number of internal degrees of freedom involved in the process, $g_{\chi}$ and $g_{\gamma}$ for $\chi$ and $\gamma$, respectively. Upon substitution of (\ref{eq:fgamma}), and following the same steps as in the usual thermal calculation (see, for example,~\cite{Gondolo:1990dk,GarciaGarcia:2016nhj}), we can rewrite the previous expression as
\begin{align}\notag
\dot{n}_{\chi} + 3Hn_{\chi} \;&=\;18g_{\chi}^2g_{\gamma}^2\frac{n_{\gamma}^2}{m_{\phi}^3}\\ \label{eq:nonthboltz}
&\hspace{-30pt}  \times \int_0^{m_{\phi}^2} ds\,\sqrt{s}\, \sigma(s) \left[\ln\left(\tfrac{m_{\phi}+\sqrt{m_{\phi}^2-s}}{\sqrt{s}}\right)-\tfrac{\sqrt{m_{\phi}^2-s}}{m_{\phi}}\right]\\
&\equiv\; \langle \sigma v\rangle_{\rm NT}\,n_{\gamma}^2\,,
\end{align} 
where the effective nonthermal cross section $\langle \sigma v\rangle_{\rm NT}$ is defined by the expression above, in analogy with the thermal case.

\subsection{Thermalization of $\gamma$}

As reheating progresses, number-conserving and violating interactions eventually bring the plasma into kinetic and chemical equilibrium. For processes mediated by a gauge interaction with coupling $\alpha$, inelastic $2\rightarrow 3$ splittings at small angles become effective at early times and rapidly populate the relativistic plasma with ``soft'' ($p\ll m_{\phi}$) particles. In turn, these (overpopulated) soft products eventually produce a fully thermalized background, that is capable of cooling down the more energetic decay products. A systematic tracking of the phase-space distributions of the hard and soft sectors reveals that the complete depletion of the hard nonthermal sector is achieved at~\cite{Harigaya:2013vwa,Harigaya:2014waa,Mukaida:2015ria}
\begin{equation}\label{eq:vth}
v_{\rm th} \simeq \alpha^{-16/5}\left(\frac{\Gamma_{\phi}m_{\phi}^2}{M_P^3}\right)^{2/5}\,,
\end{equation}
which occurs well before the end of reheating given that $\Gamma_{\phi}\ll M_P^3\alpha^8/m_{\phi}^2$, this being the case for Planck-suppressed decays and SM gauge interactions ($\alpha \sim 10^{-2}$). Note in this case that $v_{\rm th} \simeq 0.7 (|y|/10^{-5})^{4/5}$. It is also worth noting that the temperature associated with the timescale (\ref{eq:vth}), namely,
\beq
T_{\rm th} \simeq \alpha^{4/5}m_{\phi}\left(\frac{24}{\pi^2 g_{\rm reh}}\right)^{1/4}\left(\frac{\Gamma_{\phi}M_P^2}{m_{\phi}^3}\right)^{2/5}\,,
\eeq
is smaller by a factor of $\sim \alpha^{4/5}(|y|M_P/m_{\phi})^{3/10}$ than the maximum temperature $T_{\rm max}\simeq |y|^{-1/2}T_{\rm reh}$ assuming instantaneous thermalization.

Let us now go back to (\ref{eq:nonthboltz}) and note that all the time dependence is contained in the nonthermal radiation number density $n_{\gamma}$. For simplicity let us assume that the transition from the nonthermal distribution $f_{\gamma}$ to its thermal form occurs suddenly at the time (\ref{eq:vth}). 
We can then integrate with respect to time to obtain the DM number density at $v_{\rm th}$,
\begin{align}\notag
n_{\chi}(v_{\rm th})\; =\; \frac{ \langle \sigma v\rangle_{\rm NT}}{\Gamma_{\phi}}&  \left(\frac{\rho_{\rm end}}{m_{\phi}}\right)^2 \left(\frac{a_{\rm end}}{a_{\rm th}}\right)^3\\ \label{eq:solint}
&\hspace{-15pt} \times  \int_0^{v_{\rm th}} du\ (1-e^{-u})^2 \left(\frac{a_{\rm end}}{a(u)}\right)^3\,.
\end{align}

Well before the end of reheating, when the equation-of-state parameter $w=p/\rho\simeq 0$, the scale factor evolves as~\cite{EGNOP}
\beq\label{eq:scalef}
\frac{a(v)}{a_{\rm end}}\simeq \left(1+\frac{v}{A}\right)^{2/3}\,,
\eeq
where
\beq
A \equiv \frac{\Gamma_{\phi}}{m_{\phi}}\left(\frac{3}{4}\frac{\rho_{\rm end}}{m^2M_P^2}\right)^{-1/2} \simeq \mathcal{O}(1)\frac{\Gamma_{\phi}}{m_{\phi}}\,.
\eeq
Here the $\mathcal{O}(1)$ factor in the second equality is approximately equal to 2.8 for Starobinsky inflation, and to 1.3 for
a quadratic potential. Since $v_{\rm max}\sim A\ll v_{\rm th}\ll v_{\rm reh}$, the integral in (\ref{eq:solint}) can be evaluated to give
\beq\label{eq:npsiapp}
n_{\chi}(v_{\rm th})\;\simeq\; \frac{16 \Gamma_{\phi}^3 M_P^4 }{9 m_{\phi}^2 v_{\rm th}}\langle \sigma v\rangle_{\rm NT}\,,
\eeq
with $v_{\rm th}$ given by (\ref{eq:vth}). After the onset of thermal equilibrium, Eq.~(\ref{eq:nonthboltz}) becomes 
\beq
\dot{n}_{\chi}+3Hn_{\chi} = \langle \sigma v\rangle_{\rm T} (n_{\gamma}^{\rm T})^2\,,
\eeq
where $\langle \sigma v\rangle_{\rm T}$ denotes the thermally averaged production cross section, and for which (\ref{eq:npsiapp}) should be taken as an initial condition. 

Alternatively, given the linearity of the previous equation with respect to the DM number density, we can treat the thermal and nonthermal abundances as independent populations. Thus, for $v>v_{\rm th}$, the nonthermal occupation number simply redshifts. At the end of reheating, when entropy production has ceased, we can estimate the nonthermal yield as
\begin{align}\notag
Y_{\chi}^{\rm NT} \;&=\; \frac{n_{\chi}(v_{\rm th})}{n_{\gamma}^{\rm T}(v_{\rm reh})} \left(\frac{a_{\rm th}}{a_{\rm reh}}\right)^{3}\\ \label{eq:Yfinal}
& \simeq\; \frac{2 \pi ^{7/2}}{9 \zeta (3)}\left(\frac{2g_{\rm reh} }{5} \right)^{3/4} \frac{\Gamma_{\phi}^{3/2}  M_P^{5/2} }{m_{\phi}^2 }\,v_{\rm th} \langle \sigma v\rangle_{\rm NT}\,.
\end{align}
For $t>t_{\rm reh}$, this quantity is conserved (up to changes in the number of relativistic degrees of freedom) assuming that the universe has a standard thermal history after reheating.\footnote{For a nonstandard thermal history with late entropy production the final yield is reduced by the ratio of entropies before and after injection~\cite{Ellis:2017jcp,Kane:2015jia,Waldstein:2016blt}.}

\subsection{Comparison of thermal vs.~nonthermal yield}

With the previous results at hand, we can estimate the ratio of the nonthermal to the thermal abundance if we, for simplicity, assume that for a given model the total scattering cross section can be written as
\beq\label{eq:sigman}
\sigma(s) = \lambda c_n\frac{s^{n/2}}{M^{n+2}}\,,
\eeq
where $M$ is a mass scale relevant to the model, $\lambda$ is the dimensionless coupling and $c_n = {32\zeta(3)^2 }/{2^{4+n}\pi\Gamma(2+\frac{n}{2})\Gamma(3+\frac{n}{2})}$. This convoluted numerical factor is chosen so that, assuming Maxwell-Boltzmann-distributed progenitors~\cite{Gondolo:1990dk}
\begin{align}
\frac{\langle \sigma v\rangle_{\rm T}}{g_{\chi}^2g_{\gamma}^2} \;&=\; \frac{1}{\zeta(3)^2(2T)^5}\int_0^{\infty}ds\,s^{3/2}\sigma(s)K_1\left(\sqrt{s}/T\right)\\
&=\; \frac{\lambda T^n}{\pi M^{n+2}}\,. 
\end{align}
In turn, 
\beq\label{eq:fn}
\frac{\langle \sigma v\rangle_{\rm NT}}{g_{\chi}^2g_{\gamma}^2} = f(n)\,\frac{\lambda m_{\phi}^n}{M^{n+2}}\,,
\eeq
where
\beq\label{eq:fn2}
f(n)= \frac{36\,\zeta(3)^2\, \Gamma(\frac{n+5}{2})}{\sqrt{\pi}\,2^n (n+3)^2 \,\Gamma(2+\frac{n}{2}) \,[\Gamma(3+\frac{n}{2})]^2}\,.
\eeq
 Analogous expressions for $f(n)$ may be written if we assume the correct Fermi-Dirac or Bose-Einstein thermal distributions for $\gamma$. At $n=0$ the difference between these expressions and (\ref{eq:fn2})  is $\lesssim 31\%$, and decreases exponentially for larger $n$, it being $\lesssim 3\%$ for $n=6$. This correction is negligible compared to the dependence on $n$ and we will disregard it for the present discussion, but not in the detailed calculations of the following section.
%
 %

Substitution of (\ref{eq:fn}) into (\ref{eq:Yfinal}) gives
\beq
Y_{\chi}^{\rm NT} \;\simeq \;   f(n)  \frac{\pi ^5  g_{\chi}^2g_{\gamma}^2 g_{\rm reh}^{3/2} \lambda M_P \, m_{\phi}^{n-2} T_{\rm reh}^3 }{ 45 \sqrt{10} \zeta (3) M^{n+2}}\,v_{\rm th}\,.
\eeq 
Note that $Y_{\chi}^{\rm NT}\sim y^{19/5}$. For comparison, the purely thermal yield at the end of reheating is given by~\cite{Garcia:2017tuj}
\begin{align}\notag
Y_{\chi}^{\rm T}\;&\simeq\; \frac{96\zeta(3)g_{\chi}^2g_{\gamma}^2\lambda M_PT_{\rm reh}^7}{\sqrt{40}g_{\rm reh}^{1/2}\pi^4M^{n+2}}\\
&\qquad \times 
\begin{cases}
\frac{1}{n-6}(T_{\rm max}^{n-6}-T_{\rm reh}^{n-6})\,, & n\neq 6\\
\ln\left(\frac{T_{\rm max}}{T_{\rm reh}}\right)\,, & n=6
\end{cases}
\,,
\end{align}
which implies that for $n<6$, $Y_{\chi}^{\rm T}\sim y^{n+1}$. Note that this scaling suggests that $Y_{\chi}^{\rm NT}$ is a steeper function of $y$ than $Y_{\chi}^{\rm T}$ for $n<14/5\simeq 2.8$, and vice versa for $n>14/5$.

\begin{figure}[t!]
\begin{center}
{\includegraphics[width=1.0\columnwidth]{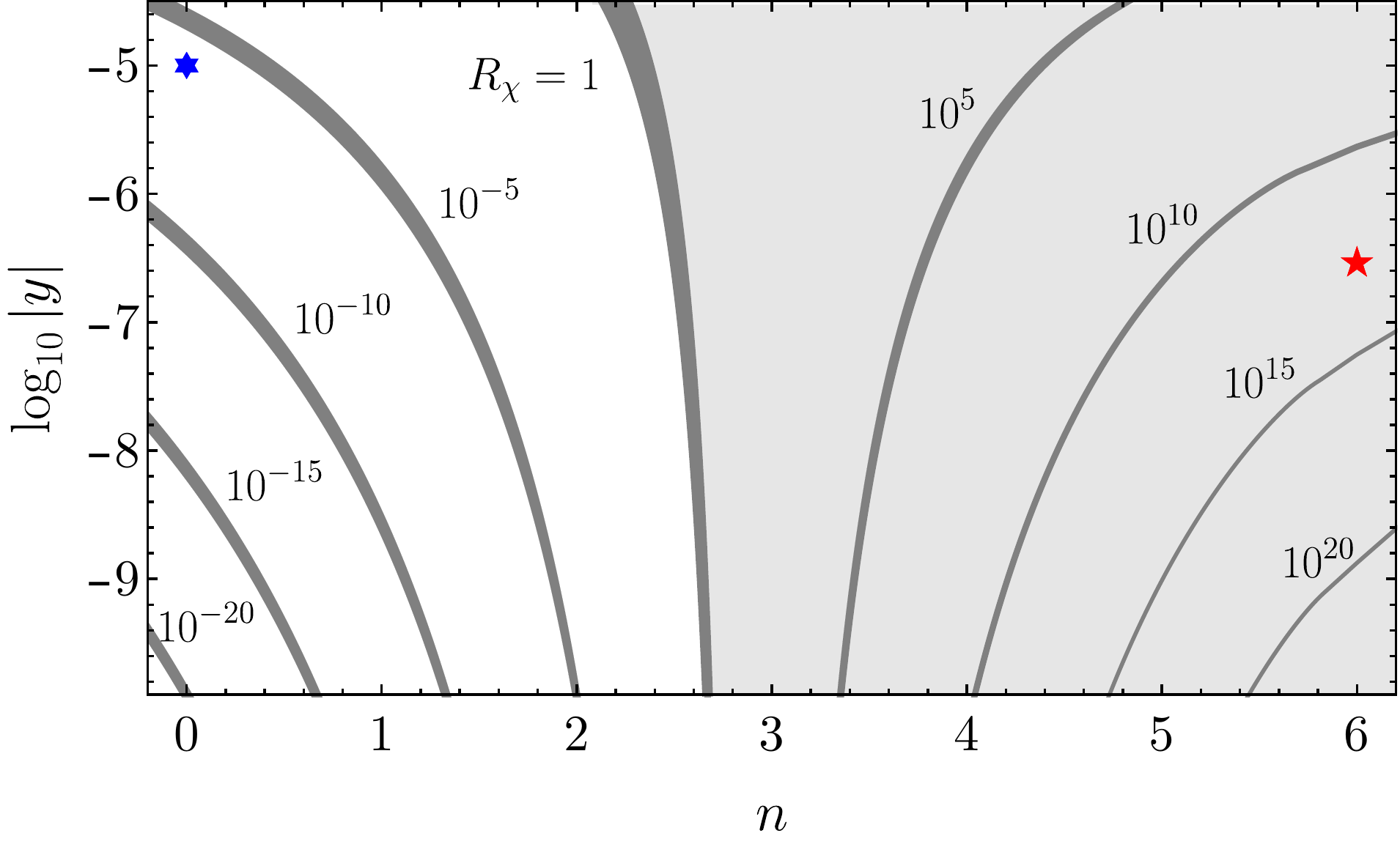}}
\vspace{-1.5em} \caption{
Contours of the nonthermal to thermal yield ratio at the end of reheating, $R_{\chi}=Y_{\chi}^{\rm NT}/Y_{\chi}^{\rm T}$, in the $(n,|y|)$ plane. The parameter $n$ determines the energy dependence of the scattering cross section, $\sigma(E)\propto E^{n}$, and $y$ is the effective dimensionless coupling that parametrizes the decay rate of the inflaton, $\Gamma_{\phi}=|y|^2 m_{\phi}/8\pi$. Thermalization is assumed to proceed via the strong gauge interaction, including one-loop $y$-dependent radiative corrections. In the shaded region nonequilibrium particle production is nonthermally dominated. The width of the level curves corresponds to the difference in gauge coupling running and in degrees of freedom between the Standard Model and its minimal supersymmetric extension, $\alpha=\alpha_3^{\rm SM}(T_{\rm reh}),\alpha_3^{\rm MSSM}(T_{\rm reh})$. The blue six-pointed star signals the upper bound for light gravitino production, while the red five-pointed star corresponds to the observed DM abundance for the EeV gravitino.}
\label{fig:regions}
\end{center}
 \end{figure}

 We can write the ratio of the nonthermal yield to the thermal one, $R_{\chi}\equiv Y_{\chi}^{\rm NT}/Y_{\chi}^{\rm T}$; for $n<6$ it has the form
\begin{align}\notag 
R_{\chi} \;\simeq\; 0.2\,(6-n)&f(n)\,  |y|^{\frac{14}{5}-n } \frac{ g_{\rm reh}^{3/2}  }{\alpha^{16/5}}\\ \label{eq:Rchi}
& \times\left(\frac{8 \pi ^4 g_{\rm reh}}{5}\right)^{n /4} \left(\frac{m_{\phi}}{M_P}\right)^{\frac{n }{2}+\frac{1}{5}}\,.
\end{align}
By itself this expression is rather obscure, which is why we show selected level curves for $R_{\chi}$ in the $(n,|y|)$ plane in Fig.~\ref{fig:regions}. For definiteness, we assume thermalization through the strong interaction, with a $y$-dependent running coupling, $m_{\phi}=3\times 10^{13}\,{\rm GeV}$ as determined from the amplitude of the scalar power spectrum~\cite{Ellis:2015pla}, and we consider both a pure Standard Model spectrum, or a Minimal Supersymmetric Standard Model (MSSM)-like spectrum; the difference is accounted for by the width of the contours. {\em In  both scenarios, we arrive to the conclusion that the DM abundance produced before thermalization is the dominant constituent for $n>2$} (assuming integer $n$). Note that the expression (\ref{eq:sigman}) is an oversimplification, as it assumes that the same production channels are available before and after the plasma reaches thermal equilibrium. Also note that the BBN bound $T_{\rm reh}\gg 1\,{\rm MeV}$, which corresponds to $y\gg10^{-13}$, is not saturated in the domain of the parameter space shown in Fig.~\ref{fig:regions}.

\section{Application to Two Models}
\label{sec:apps}

We now consider two applications of the formalism developed in the previous section. We will first revisit the nonthermal production of light gravitinos, and we will recover the results of~\cite{EGNOP} in a simpler way. We then study nonthermal production of an EeV ($10^{18}\,{\rm eV}$) gravitino~\cite{bcdm,DMO,Dudas:2018npp,Dudas:2017kfz}. We will not only focus on the final abundances, but we will also track numerically the evolution of the thermal and the nonthermal abundances during reheating, with the help of Eqs.~(\ref{eq:nonthboltz}) and (\ref{eq:vth}).

In order to avoid discontinuities in the numerical solution, we include the effect of the depletion of the nonthermal distribution and the population of its thermal counterpart for $v<v_{\rm th}$. Following~\cite{Harigaya:2013vwa,Harigaya:2014waa,Mukaida:2015ria}, the depletion of the nonthermal $\gamma$ population can be accounted for by substituting
\beq\label{eq:ngammacorr}
n_{\gamma} \;\rightarrow\; \left[1-2\sqrt{2}\left(\frac{k_{\rm split}}{m_{\phi}}\right)^{3/2} \right]\,n_{\gamma}\,,
\eeq
where
\beq
k_{\rm split} \;\simeq\; \frac{\alpha^{16}M_P^6 v^5}{2 m_{\phi}^3 \Gamma_{\phi}^2}
\eeq
denotes the splitting momentum below which a particle can deposit an order one fraction of its energy and participate in the thermal plasma. In turn, the soft component of the plasma can be considered to be partially thermalized for $v\lesssim v_{\rm th}$, and an effective temperature can be associated with it. This effective temperature is lower than the naive estimate assuming instantaneous thermalization ($v_{\rm th}\rightarrow v_{\rm max}$), and it can be used to determine the corresponding thermal number density $n_{\gamma}^{\rm T}$. We do not present here the explicit form of this effective temperature,\footnote{Explicit expressions for the effective temperature of the soft-$\gamma$ sector can be found in~\cite{Harigaya:2013vwa,Mukaida:2015ria}.} since the two corrections mentioned above are negligible, at least compared to the correction to the time dependence of the scale factor (\ref{eq:scalef}) arising from the continuous transition from matter to radiation domination around $v_{\rm reh}$.

\subsection{Light gravitino}

Gravitino production in weak scale supersymmetry breaking models has been extensively studied in the literature, the spin-3/2 particle being the poster child for the cosmological moduli problem~\cite{ehnos,Ellis:1986zt,Moroi:1994rs,Endo:2006zj,Nakamura:2006uc,Dine:2006ii,Nakayama:2012hy,Jeong:2012en,nos,kl,ekn, Juszkiewicz:gg,mmy,Kawasaki:1994af,Moroi:1995fs,enor,Giudice:1999am,bbb,Pradler:2006qh,ps2,rs,Nilles:2001ry,Nilles:2001fg,EGNOP,Ema:2016oxl,Hook:2018sai}. The production cross section is simply proportional to $M_P^{-2}$, and it is dependent on the instantaneous temperature only through radiative corrections to the gauge and Yukawa coupling. In the presence of a thermalized plasma, the average cross section for the production of gravitinos can be written as~\cite{bbb,Pradler:2006qh,rs}
\beq
\langle \sigma v \rangle = \langle \sigma v\rangle_{\rm top} + \langle \sigma v\rangle_{\rm gauge}\,,
\eeq 
with
\beq
\langle \sigma v \rangle_{\rm top} =  1.29\,\frac{|y_t|^2}{M_P^2}\left[1+\frac{A_t^2}{3m_{3/2}^2}\right] \,,
\eeq
where $A_t$ is the top-quark supersymmetry-breaking trilinear coupling, and
\begin{align}
&\langle \sigma v \rangle_{\rm gauge}  =  \sum_{i=1}^3 \frac{3\pi^2 c_i \alpha_i}{4 \zeta(3) M_P^2} \left[1+\frac{m_{\tilde{g}_i}^2}{3m_{3/2}^2}\right]
\ln\left(\frac{k_i}{g_i}\right) \notag \\
&\quad =  \frac{26.24}{M_P^2}  \left[\left(1+0.558\,\frac{m_{1/2}^2}{m_{3/2}^2}\right)\right. \notag \\
&\qquad - \left.0.011 \left(1+3.062\,\frac{m_{1/2}^2}{m_{3/2}^2}\right) \log\left(\frac{T}{10^{10}\,{\rm GeV}}\right)\right] \, ,
\label{ck}
\end{align}
where the $m_{\tilde{g}_i}$ are the gaugino masses and the constants $c_i,k_i$ depend on the gauge group (see \cite{EGNOP} for details). The first term in the gaugino mass-dependent factors $(1+m_{\tilde{g}_i}^2/3m_{3/2}^2)$ corresponds to the production of the transversally polarized gravitino, while the second term is associated with the production of the longitudinal (Goldstino) component. We therefore have $n\simeq 0$. For thermally produced gravitinos the estimate derived from the assumption of instantaneous reheating is a good approximation, as the population produced at $T>T_{\rm reh}$ is rapidly diluted by the entropy produced in subsequent inflaton decays~\cite{Nilles:2001ry,Nilles:2001fg,EGNOP,Ema:2016oxl,Garcia:2017tuj}. The yield at low temperatures $T\ll 1\,{\rm MeV}$ can be computed to give
\beq
Y_{3/2}(T)\;\simeq\; 2.54\times 10^{-6}\,|y|\, \left(1+0.56\,\frac{m_{1/2}^2}{m_{3/2}^2}\right)\,.
\eeq
Overproduction of DM is then averted for $|y|\lesssim 10^{-5}$~\cite{EGNOP}.

Let us now for definiteness consider a specific scenario in which the inflaton decays predominantly into gauge bosons, $\phi\rightarrow g+g$, as is the case in no-scale supergravity models of inflation with a nontrivial gauge kinetic function~\cite{Endo:2006xg,Kallosh:2011qk,Ellis:2015kqa}. The prethermalization gravitino abundance can then be computed as that given by the channel $g+g\rightarrow \tilde{g}+\tilde{G}$, with amplitude~\cite{bbb}
\beq
|\mathcal{M}|^2 = \frac{16\pi\alpha}{M_P^2}|f^{abc}|^2\left(1+\frac{m_{1/2}^2}{3m_{3/2}^2}\right)\left(s+2t+2\frac{t^2}{s}\right)\,,
\eeq
where $t$ and $s$ are Mandelstam variables. The resulting nonthermal production cross section is then given by 
\beq
\langle \sigma v\rangle _{\rm NT} = \sum_{i=1}^3 \frac{16\pi\alpha_i}{M_P^2}|f^{abc}|^2\left(1+\frac{m_{\tilde{g}_i}^2}{3m_{3/2}^2}\right)\,.
\eeq
\begin{figure}[t!]
\begin{center}
{\includegraphics[width=1.0\columnwidth]{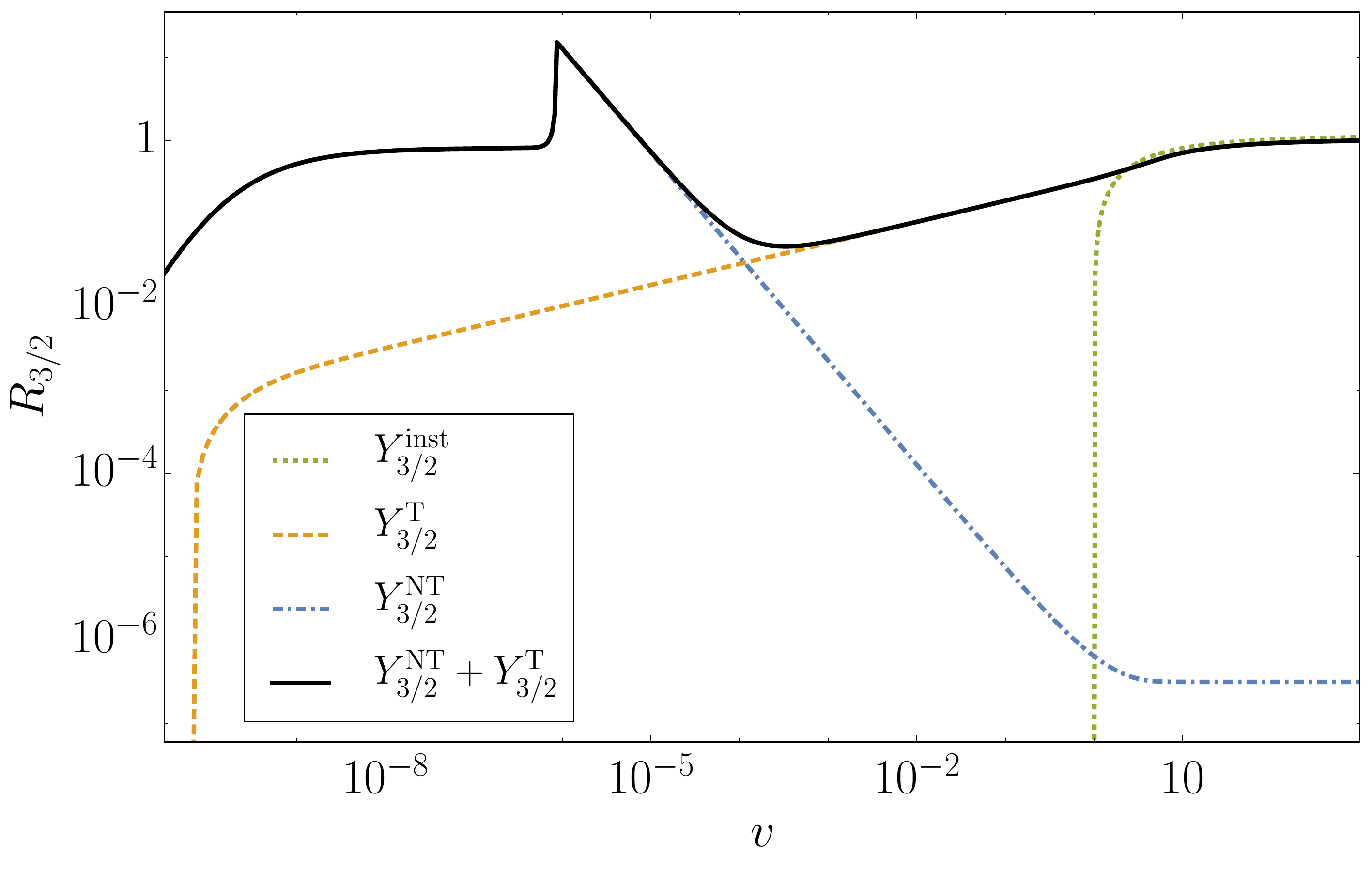}}
\vspace{-1.5em} \caption{ Ratio of the instantaneous yield to its final value after reheating assuming instantaneous thermalization, $R_{3/2}=Y_{3/2}^{X}(v)/Y_{3/2}^{\rm T}(v\gg 1)$, for the weak-scale gravitino. Here $v=\Gamma_{\phi}(t-t_{\rm end})$. Shown are the instantaneous reheating approximation (dotted green curve), the instantaneous thermalization approximation (dashed orange curve), the pure nonthermal population (dot-dashed blue curve), and the combined thermal + nonthermal result (solid black curve). The peak in the black curve is caused by the definition of $Y_{3/2}$ in terms of a single $\gamma$ degree of freedom, instead of all degrees of freedom present in the plasma. }
\label{fig:light}
\end{center}
 \end{figure}

Since the weak-scale gravitino corresponds to $n=0$ in (\ref{eq:sigman}), we expect that the early produced nonthermal abundance will be overwhelmingly drowned by the thermal population produced during the final stages of reheating; this is a known fact~\cite{EGNOP}. In Fig.~\ref{fig:light} we show the results for the numerical integration of the transport equation for $n_{3/2}$ assuming (i) instantaneous reheating and thermalization at $v=1$ (dotted green curve), (ii) instantaneous thermalization at $T_{\rm max}$ with no nonthermal production (dashed orange curve), (iii) pure nonthermal production stopping at $v_{\rm th}\simeq 10^{-6}$ (dash-dotted blue curve), and (iv) combined thermal and nonthermal production (continuous black curve). In the later case the yield is defined as $Y^{\rm NT}_{3/2}=n_{3/2}/n_{\gamma}^{\rm NT}$ for $v<v_{\rm th}$.  For simplicity we suppress the top quark contribution and the longitudinal component. The model parameters are chosen to coincide with the upper bound for DM production, $m_{\phi}\simeq 3\times 10^{13}\,{\rm GeV}$, $g_{\rm reh}=915/4$ and $|y|\simeq 2.7\times 10^{-5}$. For these, we obtain a final nonthermal population which is a factor of $\sim 3\times 10^{-7}$ less abundant than the thermally produced one, in agreement with the simple estimate (\ref{eq:Rchi}), which predicts $R_{3/2}\sim 10^{-6}$. Prethermalization effects are negligible for the light gravitino, and its production is IR dominated with respect to the reheating process, although it is UV dominated with respect to the subsequent radiation-dominated epoch.

\subsection{Heavy gravitino}

Let us consider now gravitino production in high scale supersymmetry models, where the population of the spectrum of superpartners is kinematically forbidden during reheating, since their masses are assumed to be larger than $m_{\phi}$. Only the longitudinal component of the gravitino, whose mass is significantly above the weak scale, $m_{3/2}\sim 1\,{\rm EeV}$, may be produced during reheating~\cite{bcdm,DMO,Dudas:2018npp,Dudas:2017kfz}.\footnote{This analysis is equally applicable to DM production through heavy spin-2 mediators~\cite{Bernal:2018qlk}. A recently proposed model based on an extension of the SM including a neutral gauge boson $Z'$ coupled to a set of heavy nonstandard fermions can lead to UV-dependent cross sections with $n=4,6,10$, depending on the nature of the DM particle~\cite{Bhattacharyya:2018evo}.} In this scenario, the dominant production channel is mediated by effective ${\rm SM}+{\rm SM}\rightarrow \tilde{G}+\tilde{G}$ vertices that are suppressed by the supersymmetry breaking scale $F=\sqrt{3}m_{3/2}M_P$. The total averaged production rate can be written as~\cite{bcdm}
\beq
\langle \sigma v\rangle_{\rm T} (n_{\gamma}^{\rm T})^2 = \frac{6400\pi^7}{(945)^2F^4}T^{12}\,,
\eeq
and it includes production from scalar, fermion and vector initial states. For these models $n=6$. Under the assumption of an instantaneously thermalized plasma, the final yield acquires the $T_{\rm max}$-dependent correction of the form~\cite{Garcia:2017tuj}
\begin{equation}\label{eq:Tcorrection}
Y_{3/2} \simeq 4.5\ln\left(\frac{T_{\rm max}}{T_{\rm reh}}\right) Y_{3/2,\,{\rm instant}}\,,
\end{equation}
where $Y_{3/2,\,{\rm instant}}\propto T_{\rm reh}^7 M_P/F^4$.

We now evaluate the nonthermal yield. Similar to the previous case, let us assume for simplicity an inflaton that decays into gauge bosons. The amplitude for the process $g+g\rightarrow \tilde{G}+\tilde{G}$ then has the form~\cite{bcdm}
\beq
|\mathcal{M}|^2 = -\frac{1}{32F^4}t(s+t)(3s^2+4st+4t^2)\,,
\eeq
with $t$, $s$ Mandelstam variables. Including the contributions of all 12 SM gauge bosons, the total nonthermal cross section can be computed from (\ref{eq:nonthboltz}) to be
\beq
\langle \sigma v \rangle_{\rm NT} = \frac{154 m_{\phi}^6}{5(64)^2 F^4}\,.
\eeq
If we assume that the late-time DM abundance is primarily populated by these nonthermally produced relics, we can substitute the previous expression in (\ref{eq:Yfinal}), with $g_{\rm reh}=427/4$, to obtain
\beq
Y_{3/2}^{\rm NT} \simeq 1.3\,\frac{\Gamma_{\phi}^{3/2}m^4 M_P^{5/2}}{F^4}\,v_{\rm th}\,,
\eeq
after reheating, and from (\ref{eq:omegachi}), 
\begin{align}
\Omega_{3/2}h^2 \;
&\simeq\; 2.8\times 10^{25} \,|y|^3\left(\frac{m_{\phi}}{3\times10^{13}\,{\rm GeV}}\right)^{11/2}\\
&\qquad  \times \left(\frac{0.1\,{\rm EeV}}{m_{3/2}}\right)^3 v_{\rm th}\,,
\end{align}
at $T\ll 1\,{\rm MeV}$. Substitution of (\ref{eq:vth}) assuming again thermalization through the strong interaction finally gives
\begin{align}\notag 
\Omega_{3/2}h^2 \;\simeq \; &0.11 \left(\frac{|y|}{2.9\times 10^{-7}}\right)^{19/5} \left(\frac{m_{\phi}}{3\times10^{13}\,{\rm GeV}}\right)^{67/10}\\ \label{eq:omegaheavy}
& \times\left(\frac{0.1\,{\rm EeV}}{m_{3/2}}\right)^3 \left(\frac{0.030}{\alpha_3}\right)^{16/5}\,.
\end{align}
Therefore, the observed DM relic density is obtained for $|y|\simeq 2.9\times 10^{-7}$, or $T_{\rm reh}\simeq 2\times 10^8\,{\rm GeV}$. In comparison, the assumption of instantaneous reheating and thermalization yields~\cite{DMO}
\begin{align}\notag
\Omega_{3/2}^{\rm inst}h^2 \;\simeq\; &0.11\left(\frac{0.1\,{\rm EeV}}{m_{3/2}}\right)^3\left(\frac{m_{\phi}}{3\times10^{13}\,{\rm GeV}}\right)^{7/2}\\
&\times \left(\frac{|y|}{2.9\times 10^{-5}}\right)^{7}\,,
\end{align}
which underestimates the final abundance by a factor of $\sim 10^{-14}$ given the same $y$.
\begin{figure}[t!]
\begin{center}
{\includegraphics[width=1.0\columnwidth]{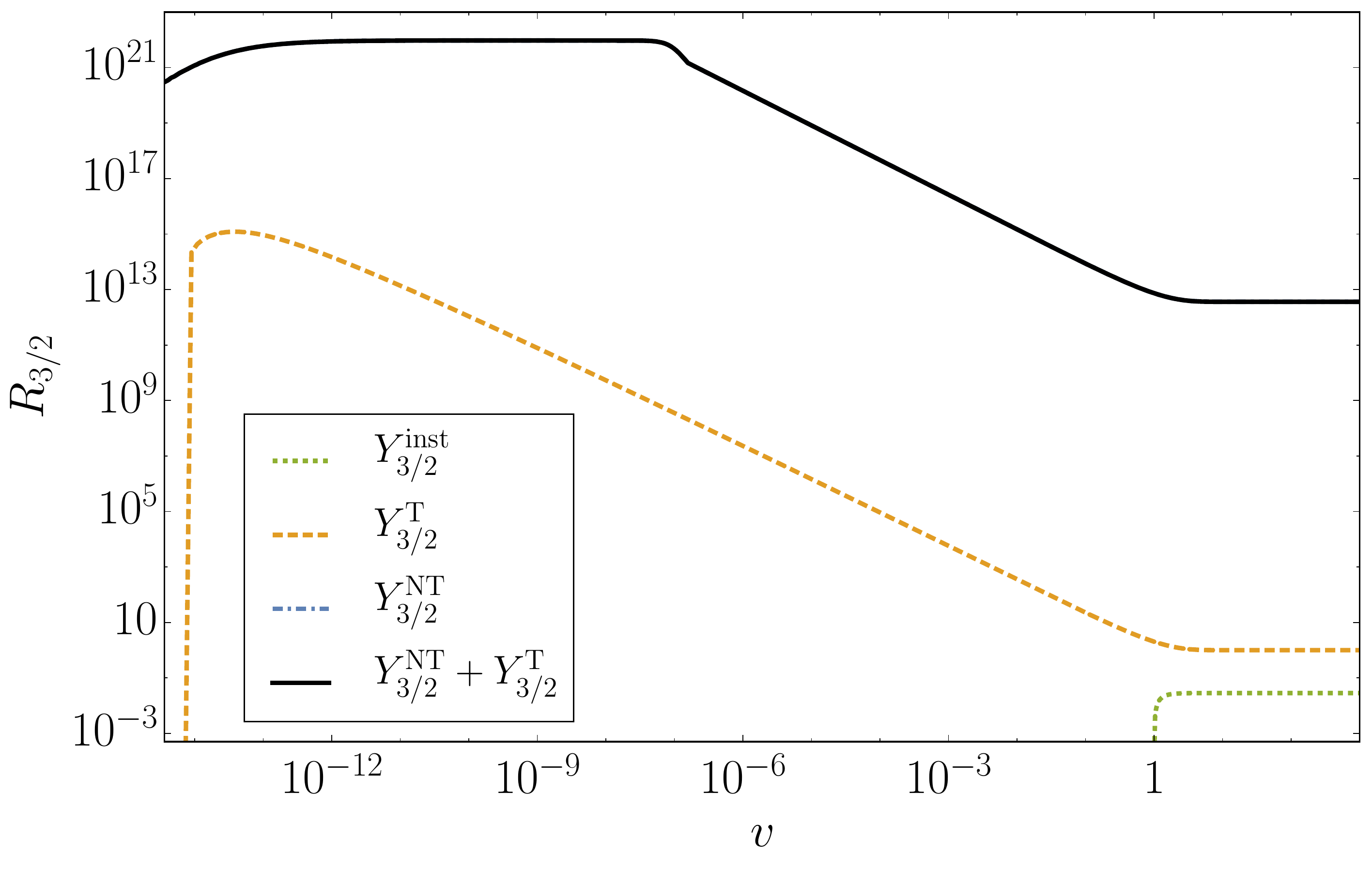}}
\vspace{-1.5em} \caption{ Ratio of the instantaneous yield to its final value after reheating assuming instantaneous thermalization, $R_{3/2}=Y_{3/2}^{X}(v)/Y_{3/2}^{\rm T}(v\gg 1)$, for the EeV gravitino. Here $v=\Gamma_{\phi}(t-t_{\rm end})$. Shown are the instantaneous reheating approximation (dotted green curve), the instantaneous thermalization approximation (dashed orange curve), the pure nonthermal population (dot-dashed blue curve), and the combined thermal + nonthermal result (solid black curve). The dot-dashed blue curve and the continuous black curve overlap for all $v$. Despite the huge enhancement for the nonthermal yield relative to the thermal yield, its maximum value is $Y_{3/2}^{\rm NT}\simeq 4\times 10^{-6} \ll 1$, consistent with the nonequilibrium assumption.}
\label{fig:heavy}
\end{center}
 \end{figure}

Figure~\ref{fig:heavy} shows the results for the numerical integration of the transport equation for $n_{3/2}$ under the same assumptions discussed in the previous section. The model parameters are chosen to yield the observed DM abundance; they coincide with their analytical estimates in (\ref{eq:omegaheavy}) within $\sim 5\%$. Here the gauge coupling running accounts only for SM degrees of freedom, with $v_{\rm th}\simeq 10^{-7}$. It is clear from the figure that, by the end of reheating, the overwhelming majority of the DM yield is produced before thermalization, with the instantaneous thermalization assumption underestimating it by a factor of $4\times 10^{12}$, and the instantaneous reheating approximation by $\sim 10^{14}$ . It is also worth noting that, despite the enormous enhancement of $n_{3/2}$ relative to the thermal result, one can check that $Y_{3/2}\ll 1$ at all times, consistent with the nonequilibrium approximation.

\section{Conclusion}
\label{sec:conc}


In this paper, we considered the production of DM from a relativistic plasma produced perturbatively at the end of inflation.
We provided a simple criterion to determine when DM production from the highly energetic, nonthermal population of the relativistic plasma provides the dominant contribution to the present DM abundance. For the production cross section $\sigma(E)\sim E^n$, if $n>2$, the nonthermal contribution dominates, whereas for $n\le 2$ the production of entropy dilutes away the nonthermal contribution. In the former case, the timescale for the onset of thermalization is crucial to determine the particle abundances at late times, due to the higher energies of the inflaton decay products compared to their thermally distributed counterparts. In the latter case, the usual  calculations assuming a thermal plasma suffice. 


We tracked the (nonthermal) phase space distributions of the relativistic plasma (produced by the perturbative decay of the inflaton) as well as the DM distribution function (sourced by the relativistic plasma) using coupled Boltzmann equations with certain simplifying assumptions. Using this analysis, we provided analytic estimates of the nonthermal yield of DM. When compared with the thermal yield, the nonthermal yield can be larger by many orders of magnitude 
(see Fig. \ref{fig:regions}). Under the assumption of gauge-mediated thermalization and Planck-suppressed inflaton decay, Eq.~(\ref{eq:Yfinal}) provides a simple analytical estimate for the nonthermal DM yield at the end of reheating. 

Our determination of the nonthermal production rate benefits from several simplifying approximations. We made use of the noninteracting decay distribution function to integrate the transport equation to determine $n_{\chi}$. However,  the relativistic plasma slowly but steadily evolves into a thermalized fluid and its constituents interact sufficiently often to create DM. For UV-dominated models, it may then be necessary to consider a more careful account of the production mechanism, 
such as the backreaction of DM production on the thermalization rate~\cite{Harigaya:2014waa}.

In terms of concrete models, we considered thermal and nonthermal creation of gravitinos in supersymmetry breaking models with a superparticle spectrum lying at the weak scale, or above the inflationary scale. In the former case, the early primordial abundance is diluted by entropy production, and the instantaneous reheating and thermalization assumption is a remarkably good estimate (as expected, since $n=0$ here). In contrast, for the EeV gravitino, the steep dependence of the scattering cross section on the momenta of the progenitors ($n=6$) leads to a relic density that is very sensitive to the prethermalization epoch; the observed abundance is obtained for a reheating temperature 2 orders of magnitude smaller than that assuming instantaneous decay at $t_{\rm reh}$. 


The derivation of the DM production rate presented here relies on four ingredients: (i) a slowly decaying, oscillating scalar field that dominates the energy budget of the universe, (ii) whose coupling to the dark sector is negligible, (iii) and which injects entropy into the universe in the form of relativistic degrees of freedom, (iv) which in turn do not thermalize instantaneously. Although these ingredients can be clearly present in the reheating epoch that follows the end of inflation, they will also generically be present during a possible moduli-dominated era. Their effect must be accounted for in addition to other ``nonthermal" phenomena~\cite{Kawasaki:1995cy,Moroi:1994rs,Acharya:2008bk,Acharya:2009zt,Dutta:2009uf,Evans:2013nka,Allahverdi:2013noa,Easther:2013nga,Blinov:2014nla,Allahverdi:2014dqa,Kane:2015jia,Georg:2016yxa,Waldstein:2016blt,Allahverdi:2017sks,Giblin:2017wlo}. Our results also open the door for consideration of nonperturbative, preheating effects in determining the DM abundance in UV sensitive scenarios, which will be pursued elsewhere. It must be noted that while nonthermal effects can impact the DM abundance, this abundance alone {\it cannot} be a model independent probe of the nonthermal processes. Additional observables, or a definite model are required to make this inference possible.

Finally, while we have focused on DM production, a similar analysis can be applied to the production of feebly interacting relativistic relics (dark radiation) through their contribution to the effective number of light relativistic species $N_{\rm eff}$ and its effect on the cosmic microwave background~\cite{Shvartsman:1969mm,Steigman:1977kc,Mangano:2001iu,Hou:2011,Rossi:2014nea,Ade:2015xua,Baumann:2015rya}. We leave this study for a future publication. 

\acknowledgments
We thank Keith Olive, Marco Peloso, and Yann Mambrini for their early interest and helpful discussions. We also thank Peter Adshead, Adrienne Erikcek, Rouzbeh Allahverdi and Scott Watson for useful conversations. M.A.~and M.G.~are supported by the U.S.~Department of Energy Grant No.~DE-SC0018216.

\bibliography{nonth}

\end{document}